\documentclass[twocolumn,amsmath,amssymb]{revtex4}

\def\gsim{ \lower .75ex \hbox{$\sim$} \llap{\raise .27ex
\hbox{$>$}} }
\def\lsim{ \lower .75ex \hbox{$\sim$} \llap{\raise .27ex
\hbox{$<$}} }

\usepackage{graphicx}
\usepackage{dcolumn}
\usepackage{bm}
\usepackage{graphicx}

\usepackage{amssymb}
\usepackage{bbm}
\def\ba{\begin{eqnarray}}
\def\ea{\end{eqnarray}}
\def\be{\begin{equation}}
\def\ee{\end{equation}}
\def\ben{\begin{equation} \nonumber}
\def\een{\end{equation}}
\def\baray{\begin{eqnarray*}}
\def\earay{\end{eqnarray*}}

\def\fid{{\dot{\phi}}}

\begin{document}

\title{A Wrinkle in Coleman - De Luccia}

\author{Adam R. Brown$^\ddagger$, Saswat Sarangi$^\ddagger$, Benjamin Shlaer$^\natural$ and Amanda Weltman$^\flat$}

\affiliation{$^{\ddagger\,\flat}$ISCAP, Columbia University, New York, NY 10027,
USA}
\affiliation{$^\natural$Physics Department, University of Colorado,
Boulder, CO 80309  USA}
\affiliation{$^\flat$Department of Mathematics and Applied Mathematics, University of Cape Town, 7700, South Africa}


\begin{abstract}

Stringy effects on vacuum transitions are shown to include surprisingly large decay rates through 
very high potential barriers.  This simple, yet counter-intuitive result will drastically modify the measure on the landscape of string vacua.
\end{abstract}

\maketitle
The decay of a false vacuum in a quantum field theory with the conventional $\frac{1}{2}(\partial_{\mu}\phi)^2$ kinetic term was first studied by Coleman\cite{coleman} via the semiclassical approximation. The effects of gravitation were later included\cite{cdl}. This work found application in the string landscape, where vacuum decay offers the mechanism by which the multitude of string vacua can be sampled \cite{landscape}. However, the action appropriate in an open string context has a richer structure than that considered in \cite{coleman, cdl}. We will show that this structure can radically modify the lifetime of certain vacua.  
To exemplify these modifications, in this letter we consider the case where open string moduli explore the string landscape.  
Branes and the charges they carry are a crucial part of stable compactifications, and the dynamics of such objects under any process of vacuum selection warrants careful study.
In the context of D3 branes probing a type IIB compactification\cite{KKLT} with warp factor $1/f(\phi)$ and potential $V(\phi)$\cite{Baumann:2006th}, the relevant action is of the Dirac-Born-Infeld (DBI) type. 
Using this
action we will discover two novel features:  that
the dominant instanton may have a ``wrinkle" (see Fig. 2), and that
raising the potential barrier can increase the tunneling rate. The net
result is that as the barrier height reaches the string scale, the decay rate becomes orders of
magnitude faster than the CDL prediction. This effect can be traced to the fact that nucleated bubbles of true vacuum can have parametrically
lower domain wall tensions than those arising from the CDL theory.
 We assume the reader is familiar with the original work by Coleman and De Luccia, which we follow closely. 
 \begin{figure}[h] 
   \centering
   \includegraphics[width=2.8in]{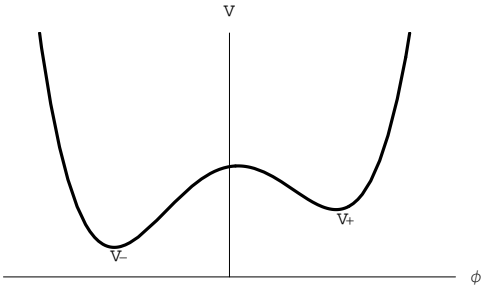} 
   \caption{A potential} 
   \label{potential}
\end{figure}

In the semi-classical approximation, the amplitude for decay of the false vacuum is calculated around a single field configuration in the path integral, namely the dominant saddle point of the Euclidian action.  The
decay rate per unit volume associated with this process is rather easily computed in Euclidean space as $\Gamma/V \sim e^{-B}$, 
where $B = \Delta S_E$ is the action difference between the bounce and false vacuum.  Coleman defines the bounce as the Euclidean solution to the equations of motion which a) asymptotically approaches the false vacuum at Euclidean infinity, b) is not a constant, and c) has smaller Euclidean action than any other solution which meets the previous two conditions.

\vspace{1mm}
{\bf Neglecting Gravity.}
It can be shown that the bounce preserves an
O(4) symmetry of our flat background.  The Euclidean DBI action is then
\be \label{eq:ignore}
S_E = 2 \pi^2 \int \rho^3 d\rho \left(\frac{1}{f(\phi)}\sqrt{1 + f(\phi) \dot{\phi}^2} -\frac{1}{f(\phi)}+ V(\phi) \right) \,\, ,
\ee
where dots represent derivatives with respect to $\rho$. 
We work in the setting of warped compactifications, where $1/f(\phi)$ denotes both the
warping and the local D3 tension.    
The unwarped case $f(\phi) = \alpha'^2$ contains the qualitative
features of open string vacuum decay, but we shall retain the warp factor $f(\phi)$ in our discussions so as to make transparent
the application of our work to more general scenarios.

We focus on $V(\phi)$ with two local minima, as in Fig. \ref{potential}.  Vacuum decay proceeds by the nucleation of bubbles of true vacuum ($V_-$) within the false vacuum ($V_+$). These bubbles will materialize quantum mechanically and then expand classically at close to the speed of light.  
Formally, Eq.(\ref{eq:ignore}) can be viewed as the action of a point particle in a potential $-V(\phi)$, albeit with an unusual kinetic term.
The O(4) symmetry of this problem is exploited by using spherical coordinates, which results in some rather unfamiliar canonical variables.  These variables can be redefined in a very simple way which keeps the O(4) symmetry manifest, while at the same time appearing Cartesian.  The cost of this choice is the addition of a friction-like term c.f. Eq.(\ref{eqn:pidot}-\ref{eqn:hdot}).  Thus, we use conjugate momentum, $\pi_\phi$, and Hamiltonian, $H$, which are defined with respect to $L^{\rm canonical}/(2\pi^2 \rho^3)$ unless explicitly stated. 
\ba
L = \frac{L^{\rm{canonical}}}{2\pi^2\rho^3} &=& \frac{1-\gamma}{f \gamma} + V \,\, , \\ 
\pi_\phi = \frac{\partial L}{\partial \fid}  &=& \fid \gamma \,\, , \\
H = \pi_\phi \fid - L  &=& \frac{1-\gamma}{f} - V\label{eqn:hamiltonian} \,\, , 
\ea
where
\ba
\gamma = \frac{1}{\sqrt{1 + f(\phi)\fid^2}} = \sqrt{1 - f(\phi) \pi_\phi^{2}} \label{gammadef}\,\, .
\ea
Notice that in Euclidean space, it is the relativistic momentum which is bounded, and not the velocity\cite{silversteintong}. 
As $\pi_\phi \to f(\phi)^{-1/2}$ the field velocity $\fid \to \pm\infty$.
The equations of motion are easily derived from the canonical equations of motion, yielding
\ba
\dot{\pi}_\phi = -\frac{\partial H}{\partial \phi} - 3 \frac{\pi_\phi}{\rho} \label{eqn:pidot} \,\, .
\ea

An important result is
\be
d H = - \frac{3}{\rho} \pi_\phi d\phi 
\label{eqn:hdot} \,\, , 
\ee
which determines the amount of non-conservation of $H$.  
As promised, the spherical measure induces friction in this simple one-dimensional system.  Since a meaningful trajectory begins and ends with zero velocity $(\gamma = 1)$, $dH$ integrates to the energy difference between the center and outside of the bubble.  (This amounts to $V_+ - V_-$ in the thin wall limit.) 

\vspace{1mm}
{\bf  Thin Wall Approximation.}
In the thin wall approximation, we require $S_1 \gg \epsilon/\mu$, where $\epsilon$ is the difference in vacuum energy, $\mu$ is the mass of $\phi$ (in either vacuum), and $S_1$ is the tension of the domain wall.  In the inverted potential, $\phi$ sits atop the true vacuum, $\phi_{-}$, for a long time before falling off and rolling quickly to the top of the false vacuum, $\phi_{+}$.  The thin wall bounce is nearly a step function.  
A first integral of the equation of motion is achieved by writing $H + \mathcal{O}(\epsilon) = E$ in Eq.(\ref{eqn:hamiltonian}), yielding
\be
\gamma(\phi,\pi_\phi) = 1 - f(\phi)\left(V(\phi) +E + \mathcal{O}(\epsilon)\right) \label{eqn:solvegamma} \,\, , 
\ee
where $E = -V_+$, the value of the potential at the false vacuum.  This can be solved for the momentum to give
\be
\pi_\phi = \sqrt{V_0(\phi)\left(2 - f(\phi) V_0(\phi)\right)}  \,\, , 
\ee
where we have absorbed $E + \mathcal{O}(\epsilon)$ by setting the minima of $V_0$ equal to zero.
From this we can write the formal solution $\phi = \phi(\rho)$
\be
\rho(\phi) = \int \frac{1 - f(\phi)V_0(\phi)}{\sqrt{V_0(\phi)\left(2 - f(\phi) V_0(\phi)\right)}}d\phi  \,\, . \label{eq:twsoln}
\ee

We may use these solutions to find the bubble nucleation rate.  We are interested in
the difference in the Euclidean action between the above bounce and the static solution of sitting atop the
false vacuum for all $\rho$.  Thus, there are two contributions to $B = \Delta S_E$: the interior volume
and the bubble wall.

\ba
B &=&  -2 \pi^2 \int_0^{\bar{\rho}} \rho^3   \epsilon d\rho  + 2\pi^2 {\bar{\rho}}^3\int_{\phi^-}^{\phi^+}\sqrt{V_0(2 - f  V_0)}d\phi  \,\, , \nonumber\\
&=& -\frac{\pi^2{\bar{\rho}}^4}{2}\epsilon + 2\pi^2{\bar{\rho}}^3S_1 \,\, ,
\ea
where $\bar{\rho}$ is the radius of the bubble.
We may integrate Eq.(\ref{eqn:hdot}) to solve for
$\bar{\rho} = 3 S_1/\epsilon$.
The tunneling rate per unit volume is given by
\be
\Gamma/ V \sim e^{-27\pi^2 S_1^4/2\epsilon^3} \,\, . 
\ee
The bubble wall tension is $S_1 = \int d\phi\sqrt{V_0(2-fV_0)}$, so the tunneling rate is large for $f V_0 \to 2$.  This tunneling rate is always greater than the Coleman - De Luccia rate in the same potential, and reduces to the CDL result in the limit $f \rightarrow 0$.  

\vspace{1mm}
 
 {\bf Discussion.}
As one might expect from a covariant Euclidean action, there is a generic multi-valuedness  
of $\phi(\rho)$.  The complete range of $\gamma$ is given by
\be
-1 \leq \gamma \leq 1 \quad\Longrightarrow \quad 0 \leq f(\phi) V_0(\phi) \leq 2 \,\, .
\ee
To borrow Lorentzian terminology, we can think of $\gamma \sim 1$ as ``non-relativistic," and $\gamma \ll 1$ as ``relativistic."
The lower bound on $fV_0$ is where the motion becomes classical, and the upper bound is where
nucleated antibrane motion becomes classical.  

It should be noted that when $fV_0$ has a range exceeding two, no continuous interpolating solutions exist.  Instead, nucleation of branes can occur via the Brown-Teitelboim (BT) mechanism\cite{Brown:1988kg}.  One
may think of the enhanced tunneling rate we find as due to the proximity of the BT threshold.  The motion of nucleated branes is classical, and thus does not incur exponential suppression.  In this sense, our tunneling scenario is a deformation of the CDL instanton toward the BT instanton.

\begin{figure}[h] 
   \centering
   \includegraphics[width=3in]{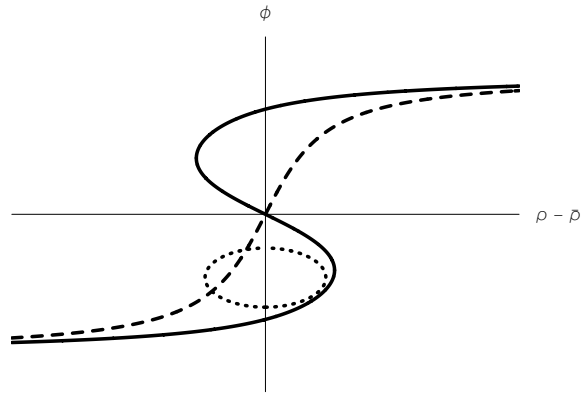} 
   \caption{A ``non-relativistic" (dashed),  ``relativistic" (solid), and supercritical  (dotted) profile for $\phi(\rho)$. Regions with
   negative slope describe the wrinkle of ``antibrane".  The supercritical case does not describe a transition, but rather brane nucleation.} 
   \label{profile}
\end{figure}

A typical thin walled solution would look something like Fig. \ref{profile}.
The wrinkle in the brane for large $f V_0$ can be thought of as the desire for the
brane to become an antibrane.  The wrinkle allows the orientation of the brane
to be reversed at the expense of surface area.  This is obviously favorable when
either $V_0$ becomes large or $1/f$ (the brane tension) becomes small.

\vspace{1mm}
{\bf Validity of the DBI action.}
The DBI action can be trusted whenever curvatures are low \cite{Leigh:1989jq}.  
In the thin wall regime, the extrinsic curvature of the solution $\phi(\rho)$ (from Eq.(\ref{eq:twsoln})) in the warped geometry is given by
\begin{eqnarray}
K(\phi)
 &=&  \frac{f(\phi)^{3/4}}{\sqrt{\alpha'}}\frac{\partial}{\partial \phi}\left(V(\phi)-\frac{1}{f(\phi)}\right)\,\, .
 \end{eqnarray}
 
Since consistency requires that $K$ be much less than the invariant string scale $1/\sqrt{\alpha'}$, we should only consider potentials $f(\phi)$ and $V(\phi)$ with limited steepness.  One consequence of this
is that we can only consider backgrounds with low AdS curvature, or equivalently, gauge theories with large 't Hooft coupling $\lambda \gg 1$.

When staying within the validity of our approximation, the bubble wall tension must still be greater than the local string scale of the two relevant vacua.
This can be exponentially small compared to the scales of the bulk through which the brane tunnels, and can thus be exponentially smaller than the standard CDL result.  Since the decay rate of the false vacuum is exponentially sensitive to the bubble wall tension $S_1$, the deviation from CDL can be
many orders of magnitude.



\vspace{1mm}

{\bf Including Gravity.}
In this section we study the decay of the false vacuum with the DBI action in the presence of gravity. We neglect a possible $R \phi^2$ term, but will return to it in a future publication \cite{bssw}. We maintain our O(4) symmetric ansatz, so the Euclidean metric is $ds^2 = d\xi^2 + \rho(\xi)^2 d\Omega_{3}^2$ where $\rho$ here plays the role of a scale factor and $\xi$ is radial coordinate. After integrating by parts, the Euclidean action is given by
 %
%
\ba
S_E &=& 2 \pi ^2 \int d\xi \rho ^3 \left(\frac{\sqrt{1 + f(\phi) \dot{\phi}^2}-1}{f(\phi)}+V(\phi) \right. \nonumber \\ \label{Sparts}
&&\left.-\frac{ M_P^2}{2} \frac{6 \left(1+\dot{\rho}^2\right)}{ \rho ^2} \right) \,\, ,
\ea
where $M_P$ is the reduced Planck mass, and dots now signify derivatives with respect to $\xi$.
The canonical Hamiltonian is thus
\ba
H^{\rm canonical} &=& 2\pi^2\rho^3 \left(   \frac{ - 1}{f(\phi ) \label{gravham}
   \sqrt{1+ f(\phi ) \dot{\phi} ^2}} + \frac{1}{f(\phi)}-V(\phi) \right. \nonumber \\
   &&\left. +\frac{3 M_P^2 \left(1-\dot{\rho}^2\right)}{\rho ^2}       \right) \,\, .
\ea
Solving the $\xi \xi$ Einstein equation is equivalent to the Hamiltonian constraint $H^{\rm canonical} = 0$, which we use to obtain
\be
\label{einsteq}
\dot{\rho} =  \pm \sqrt{1 + \frac{\rho^2}{3 M_P^2}\left( -V(\phi) + \frac{1}{f(\phi)} - \frac{1}{f(\phi) \sqrt{1 + f(\phi)\dot{\phi}^2}} \right)} \,\, .
\ee
For a transition from a positive cosmological constant, $\dot{\rho}$ starts out at $+1$ and curves over to zero at ``maximum circumference $S^3$" before becoming negative.  Increasing $\rho$ provides friction, and decreasing $\rho$ provides negative friction.

The momentum conjugate to $\phi$ is given by
\be
\pi_\phi = \gamma \dot{\phi}  \,\, .
\ee
%
%
Once again we have
\ba
\label{gamma}
\gamma &=& 1 - f(\phi) \left(V(\phi) + H^\phi(\pi_\phi,\phi)\right) \,\, , \\
d H^\phi &=& - 3\frac{\dot{\rho}}{\rho} \pi_\phi d\phi \label{eqn:hphidot} \,\, ,
\ea
where $H^\phi$ is the quantity in parentheses in Eq.(\ref{einsteq}).  
In the thin wall regime, we may solve the $\phi$ equation of motion via
\be
\xi(\phi) = \int \frac{1 - f(\phi)V_0(\phi)}{\sqrt{V_0(\phi)\left(2 - f(\phi) V_0(\phi)\right)}}d\phi  \,\, , \label{eq:twsolngrav}
\ee
where $V_0(\phi) = V(\phi) - V_+ + {\mathcal O}(\epsilon) + {\mathcal O}(S_1^2/M_P^2)$.
As before, a monotonic $\phi$ lets us absorb all friction into the effective potential $V_0(\phi)$.

An apparent complication arises in solving Eq.(\ref{einsteq}) because $H^\phi(\pi_\phi,\phi)$ is a multi-valued function of $\xi$ (whenever $\phi(\xi)$ is).  While unusual from a gauge theory point of view, a multi-valued $\phi$ is quite natural from a geometric perspective, where $r=\alpha'\phi$ is an embedding coordinate of the brane in the ten dimensional background.  
To be precise, the quantity in parentheses in Eq.(\ref{einsteq}) is a single-valued function of $\xi$, given by
\be
H^\phi(\xi) = \int dH^\phi(\pi_\phi,\phi) = -3 \int \frac{\dot{\rho}}{\rho}\frac{dS_1(\xi)}{d\xi}d\xi \label{hxi}\,\, ,
\ee
where
\be
S_1(\xi) = \int_{\phi^{-1}(\phi) \, < \, \xi} \hspace{-10mm} d\phi \sqrt{V_0(\phi)(2 - f(\phi)V_0(\phi))} \label{s1xi} \,\,.
\ee
The function $\phi^{-1}(\phi)$ is given by the l.h.s. of Eq.(\ref{eq:twsolngrav}).
Because Eqs.(\ref{einsteq}, \ref{eq:twsolngrav}, \ref{hxi}, \& \ref{s1xi}) are all coupled, their solution is difficult
to compute.

Luckily, coupling gravity to a multi-valued matter Hamiltonian does not complicate the calculation of the
decay rate, since the action is a well defined functional.  In the thin wall regime, a closed form solution is
even possible.  We demonstrate this below.

Using Eqs.(\ref{Sparts}) and (\ref{einsteq}), the Euclidean action can be written as
\ba
S_E = 4\pi^2 \int d\xi\left\{\rho^3\left(V + \frac{(\gamma  -1)^2}{2\gamma f(\phi)}\right)\right. - 3 M_P^2 \rho{\bigg \}} \label{rewrite}\,\, .
\ea
The action difference is $B[\phi] = S_E[\phi] - S_E[\phi_+]$. 
On the wall we have to circumvent the multi-valuedness of the integrand in Eq.(\ref{rewrite}) using $d\xi = \gamma/\pi_\phi d\phi\,$. This yields
\ba
B_{\rm wall} &=& 4\pi^2  \int_{\phi_-}^{\phi_+} d\phi \rho^3\frac{\gamma}{\pi_\phi}\left(V_0(\phi) + \frac{(\gamma -1)^2}{2\gamma f(\phi)} \right) \nonumber \\
&=& 2\pi^2\bar{\rho}^3S_1 \label{Bwall}\, \, ,
\ea
where the tension
\be 
S_1 = \int_{\phi_-}^{\phi_+} d\phi \sqrt{V_0(\phi)(2 - f(\phi)V_0(\phi))} \,\, , \label{gravtension}
\ee 
as before.  As pointed out below Eq.(\ref{eq:twsolngrav}), $V_0(\phi)$ receives corrections from the
back reaction of the domain wall, although their contribution to $S_1$ is negligible in the thin wall regime.

In the thin wall approximation, the interior field configuration coincides with
the true vacuum $V_-$\, , and $\gamma = 1$.   The contribution to $B$ inside the bubble will be the same as that found in \cite{cdl} in 
the non-DBI context.  Inside the bubble, $d\xi = d\rho \left( 1 - (\rho^2V_-)/(3M_P^2) \right)^{-1/2}$, and so the action can be written
\ba
S_E^{\rm inside} = -12\pi^2 M_P^2\int_{0}^{\bar{\rho}} d\rho \rho \sqrt{ 1 - \frac{\rho^2 V_-}{3M_P^2} }  \, \, ,  
\ea 
and the contribution to $B$ is 
\be
B_{\rm inside} = \left\{\frac{-12\pi^2 M_P^4}{V}\left[1 - \left(1 - \frac{\bar{\rho}^2 V}{3 M_P^2}\right)^\frac{3}{2}\right]\right\}^{V = V_-}_{V = V_+} \,\, .
\ee

Lastly, one may determine $\bar{\rho}$ by demanding that the action be stationary with respect to its variation, yielding
\be
\bar{\rho} = \frac{3 S_1M_P}{\sqrt{\epsilon^2 M_P^2 + \frac{3}{2} S_1^2 \left(V_+ + V_-\right) +\frac{9}{16} S_1^4/M_P^2 }}\,\,.
\ee
The decay rate per unit volume is then given by
\be
\Gamma/V \sim e^{-\left(B_{\rm wall} + B_{\rm inside}\right)}\,\, .
\ee
Comparing the value of $B$ in the three regions with the values in \cite{cdl}, the crucial change is in the tension of the bubble wall
$\int d\phi \sqrt{2 V_0} \to S_1 = \int d\phi \sqrt{V_0(\phi)(2 - f(\phi)V_0(\phi))}$. We have already seen this phenomenon in the nongravitational case.  In the ``nonrelativistic" limit of the 
action, i.e. $fV_0 \ll 1$, one recovers the usual bubble wall tension. However, when $fV_0 \to 2$ over much of the potential, the tension is drastically reduced, as is manifest in Eq.(\ref{gravtension}). This leads to an enhancement in the tunneling rate, which will radically modify the measure on the open string landscape. In forthcoming work we will extend this result to more explicit models \cite{bssw}. 

Recent progress in string compactification has led to a rather dramatic new framework for understanding the origins of our vacuum.  Within this framework, the significance of branes and their charges cannot be overstated.  It is therefore crucial to understand the role open strings play in the
dynamics of vacuum selection.
The principal formalism at our disposal is the Coleman - De Luccia instanton, and what we have demonstrated is  a significant deviation from their result, despite remaining within the semi-classical approximation.  
For an alternative approach leading to effects beyond CDL, (e.g. due to resonance tunneling among the multitude of vacua) see \cite{HenryTye:2006tg}. 
\vspace{1mm}

{\bf Acknowledgments.}
We would like to thank Shanta de Alwis, Puneet Batra, Oliver DeWolfe, Brian Greene, Dan Kabat, Louis Leblond, Jeff Murugan, Sarah Shandera and Henry Tye for useful conversations. The work of A.B. is supported by a NSF Graduate Research Fellowship, S.S. by DE-FG02-92ER40699, B.S. by DOE grant DE-FG02-91-ER-40672 \& A.W. by a NASA Graduate Student Research Fellowship grant NNG05G024H.  B.S. would also like to thank the Perimeter Institute for hospitality while this research was completed.

\end{document}